\documentclass[a4paper]{mem_sissa}
\usepackage{natbib}
\usepackage{graphicx}
\usepackage[a4paper]{hyperref}
\begin{document}
   \title{Variable Stars in the LMC:\\ the Photometric Catalogue
%\thanks{this is a place for a title footnote}
}

   \author{M. Maio \inst{1}, G. Clementini \inst{1}, A. Bragaglia \inst{1},
   \\
          E. Carretta \inst{2}, R. Gratton \inst{2} \and L. Di Fabrizio \inst{3} \fnmsep }
%%%%%\thanks{this is a place for placing a footnote in the author field }

   \offprints{M. Maio}
\mail{via Ranzani 1, 40127 Bologna }

   \institute{INAF - Osservatorio Astronomico di Bologna,
via Ranzani 1, 40127 Bologna, Italy \\ 
              \and  INAF - Osservatorio Astronomico di Padova, Vicolo
	      dell'Osservatorio 5, 35122 Padova,
 Italy\\
              \and INAF - Centro Galileo Galilei \& Telescopio Nazionale Galileo, PO Box
	      565, 38700 Santa Cruz de La Palma, Spain\\
             }

   \abstract{
   New B, V, I photometry was obtained for a sample of 152
   variables (125 RR Lyrae's, 4 anomalous Cepheids, 11 classical Cepheids, 11
   eclipsing binaries and a $\delta$ Scuti star) in two regions near the bar of
   the Large Magellanic Cloud (LMC). We derived complete and well sampled B and V
   light curves, very accurate periods, epochs of maximum light, amplitudes,
   intensity- and magnitude-averaged apparent luminosities for about 95\% of the
   RR Lyrae stars. All these informations are contained in the photometric
   catalogue. Metallicities from the Fourier decomposition of the V light curves
   were derived for 29 of the ab type RR Lyrae's and are compared with the
   $\Delta$S metal abundances from low resolution spectra obtained with the Very
   Large Telescope (VLT).  
   }
   \authorrunning{M. Maio et al.}
   \titlerunning{Variable Stars in the LMC: the Photometric Catalogue}
   \maketitle
%
%________________________________________________________________

\section{Introduction}

%                                     Two column figure (place early!)
%______________________________________________ Gamma_1 (lg rho, lg e)
   \begin{figure*}
   \centering
   %\resizebox{\hsize}{!}{
   \includegraphics[width=14cm]{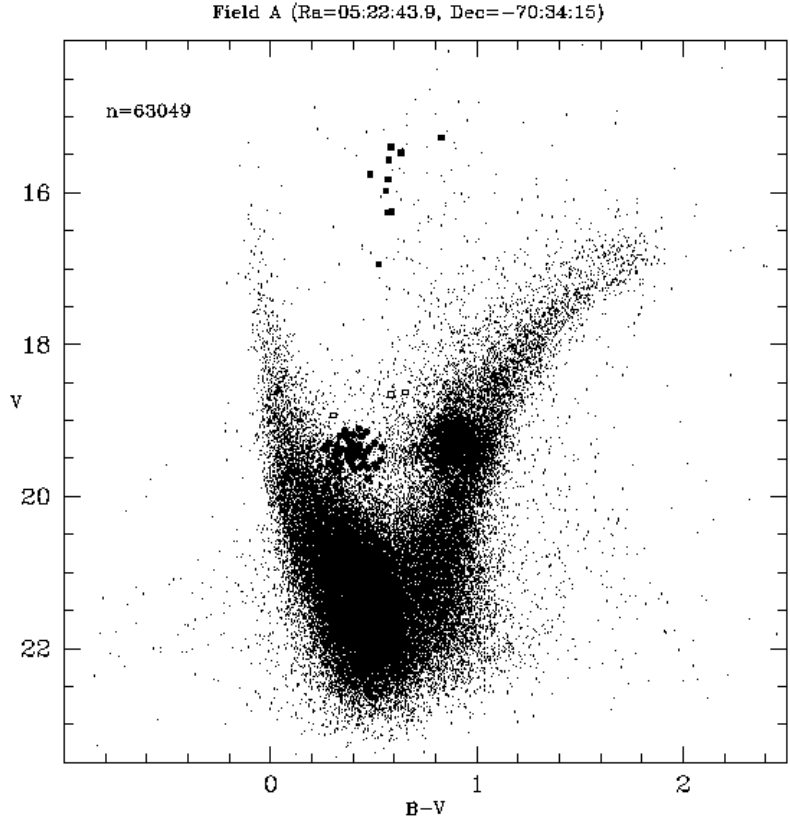}%}
   \caption{V {\em vs} B - V color - magnitude diagram of Field A from the
   ALLFRAME reduction of the 2001 data (Clementini et al. 2003). 
   Different symbols mark the variables
   identified in this field (RR Lyrae stars: filled circles; anomalous Cepheids:
   open squares; Cepheids: filled squares; binaries: filled triangles; crosses:
   $\delta$ Scuti).}
              \label{fig1}
    \end{figure*}
%______________________________________________________________

A large number of variables of various types have been
detected in the LMC by the MACHO (Alcock et al. 1996) and OGLE (Udalski et al.
2000) microlensing experiments. Discovery of more than 7900 RR Lyrae's in the
bar of the LMC is reported by Alcock et al. (1996), as well as of a large number
of Cepheids and eclipsing binaries. RR Lyrae's and Cepheids are of particular
interest since they are primary distance indicators for the LMC and for 
Local Group galaxies, in general.

Notwithstanding the large amount of observational data collected for the
variables in the LMC by the microlensing experiments, the nonstandard
photometric passbands used by some of these missions (MACHO) and the fact that RR
Lyrae's are close to the limiting magnitude of these surveys set constraints
on the use of those photometric databases.
Besides, because in these experiments variables have mainly been observed in I
and to a lesser extent in V and B, this makes more difficult the comparison
with most of the Galactic sample observations which generally are in B and V,
instead.

\section{Observations and reductions}

Time series B, V, and I exposures of two 13$\prime$ $\times$ 13$\prime$ fields
(hereinafter
called Field A and B) close to the bar of the LMC and contained in Field \#6
and \#13 of the MACHO microlensing experiment (Alcock et al. 1996) were
obtained with the 1.54 m Danish telescope at ESO (La Silla, Chile) in two
observing runs respectively in January 1999 and 2001. Field A also has a 40\%
overlap with the OGLE-II field LMC\_SC21 (Udalski et al. 2000). 
The full data-set consists of 72, 41 and 15 frames in V, B, and I respectively.
Reduction and analysis of the 1999 photometry were done using the package
DoPHOT (Schechter, Mateo \& Saha 1993), while photometric reductions of the 2001
data were done using the packages DAOPHOT/ALLSTAR II (Stetson 1998)
and ALLFRAME (Stetson 1994).
%
%                                                
%----------------------------------------------------------- 
   \begin{figure*}
   \centering
   \resizebox{\hsize}{!}{\includegraphics{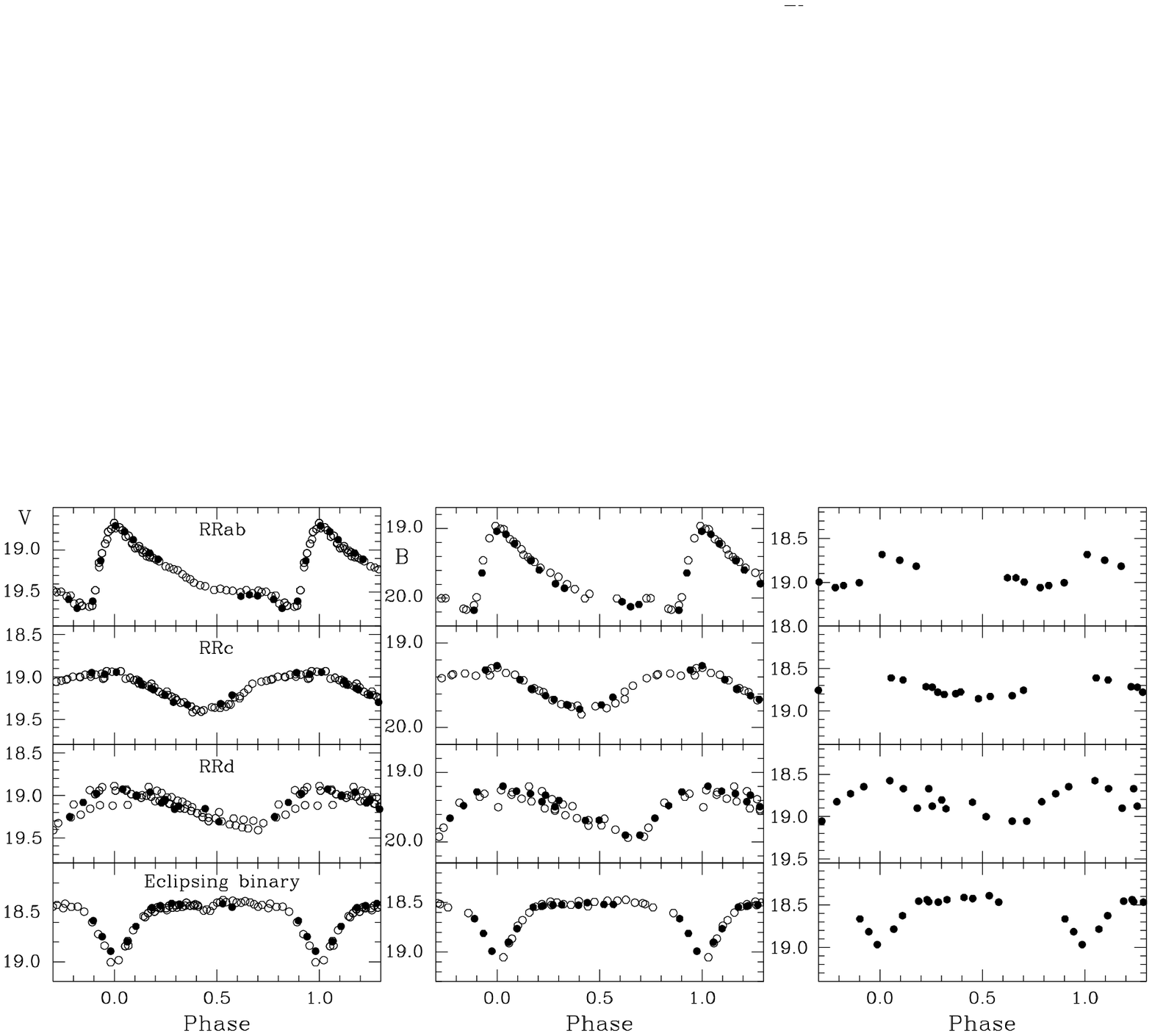}}
     \caption{Some examples of our light curves. Open circles are from
     the 1999 observations, filled dots are the new 2001 data points.
               }  
        \label{fig2}
    \end{figure*}
%
%______________________________________________________________
%
%  
%

\section {Data Analysis}

Variable stars were identified on the 1999 time series data-set using the
program VARFIND (private software by P. Montegriffo), and then counteridentified
in the 2001 data-set. VARFIND displays the scatter diagram of the average
measurements, from which candidate variables are identified due to their large
standard deviations. 
Each candidate variable identified with VARFIND was then checked for variability
using the program GRATIS (GRaphycal Analyzer TIme Series), a software
developed at Bologna Observatory by P. Montegriffo, G. Clementini and L. Di
Fabrizio, directly interfaced to VARFIND, which performs a period search on the
data using both the Lomb periodogram (Lomb 1976, Scargle 1982) and a best-fit of
the data with a truncated Fourier series (Barning 1963). About two thousand
objects were
checked for variability in both field A and B. Among them we detected
152 certain variables and in addiction 7 candidates of unknown type. In the
sample, there are 125 RR Lyrae stars (77 RRab, 38 RRc, 10 RRd), 4 anomalous
Cepheids, 11 classical Cepheids, 11 eclipsing binaries and one $\delta$ Scuti
star.
Our identification of the RRab variables in Field B should be complete, while
identification may be less complete in Field A, because of the larger crowding
conditions. Moreover some RRc's may have escaped
detection in both fields, due to their smaller amplitudes. 
Figure \ref{fig1} shows the location of the variable stars on the HR
diagram of Field A; variables are plotted according to their intensity-average
magnitudes and colors, and with the different symbols corresponding to the
various types.

Thanks to the two years time base line spanned by our observations,
we were able to: 
\begin{description}
   \item[a)]derive periods and epochs for all the 152 variables with a precision
   better than the fourth decimal place, depending on the light curve data
   sampling (in the best cases 72 V, 41 B and 14 I);
   \item[b)]solve aliasing problems and check the actual variability of some
   candidate variables detected in 1999;
   \item[c)]identify the Blazhko modulation of the light curves in about 15\% of
   the total sample of our RR Lyrae stars;
   \item[d)]derive complete and well sampled B and V light curves for about 95\%
   of the RR Lyrae stars.
\end{description}

GRATIS allows to perform also a search for multiple periodicities and was run
on the data of the 10 double-mode pulsators falling in our fields, one of which
newly discovered.
In figure \ref{fig2} we show the multicolor light curves for an RRab, an RRc,
an RRd and an eclipsing binary in our sample.
The period distribution of the single-mode RR Lyrae's 
shows two peaks, corresponding to the {\em ab} (77 objects) and
{\em c} (38 objects) type pulsators. The mean periods are 0.$^d$580
($\sigma$=0.064) and  0.$^d$327 ($\sigma$=0.047) for {\em ab} and {\em c}
pulsators, respectively.
The preferred period of the RRab's falls between the periods of the Galactic
globular clusters of Oosterhoff (1939) type I (OoI) and II (OoII), and actually
closer to the OoI type.
The overlap in the transition region between {\em ab} and {\em c} type is small
(6 objects) and the transition period occurs at P$_{\rm tr} \sim 0.^d$40.
B and V amplitudes were calculated as the difference between maximum and minimum
of the best fitting models for all the RR Lyrae's having full coverage of the
light curve. These amplitudes have been used together with the derived periods
to build-up period-amplitude diagrams. We compared the derived period-amplitude
distributions with relations defined by the RRab's in the globular
clusters M3, M15 and $\omega$ Centauri. RR Lyrae's in Field B seem to follow
better the amplitude-period relations of the variables in M3 and to belong to
OoI type. Variables in Field A, instead, have pulsational properties more
intermediate between the two Oosterhoff types. 
We derived also metallicities and absolute magnitudes from the parameters of the
Fourier decomposition of the V light curves; we applied Jurcsik \& Kov\'acs
(1996) and Kov\'acs \& Walker (2001) techniques to 29 RRab's that satisfy the
completeness and regularity criteria defined by Jurcsik \& Kov\'acs
(1996). The derived average metallicity is [Fe/H]=$-1.27$ dex ($\sigma$=0.35; 
on Jurcsik 1995 metallicity scale). 
This value is
in good agreement with our previous estimate of [Fe/H]=$-1.49\pm0.11$ dex
($\sigma$=0.28) (Bragaglia et al. 2001) determined applying the $\Delta$S
method (Preston 1959) to 6 RRd's, once differences in the metallicity scales are
taken into account.
Metallicities obtained in 2001 for 101 of the RR Lyrae stars in our sample, from
a spectroscopic survey with FORS at the VLT, are again in agreement with the
previous results, giving an average metal abundance of [Fe/H]=$-1.48\pm0.03$ dex
($\sigma$=0.29; Gratton et al. 2003). 

\bibliographystyle{aa}

\end{document}